\documentclass[acmlarge,screen,11pt,nonacm]{acmart}
\setcopyright{none}	
\AtBeginDocument{%
  }

\usepackage{subcaption}  
\usepackage{longtable}  
\usepackage{mathpartir}  
\usepackage[normalem]{ulem}  
\usepackage{caption} 
\usepackage{multirow}
\usepackage{xcolor}
\usepackage[Algorithm]{algorithm}
\usepackage{algorithmic}
\usepackage{array}
\usepackage{graphicx}

\usepackage{ragged2e}
\newcolumntype{L}[1]{>{\RaggedRight\arraybackslash}m{#1}}

\hypersetup{ 
	colorlinks=true,
	urlcolor=blue,
	linkcolor=black,
	citecolor=black
}

\begin{document}

\title{SafeLLM4SE: Statistical Evaluation and Reporting for LLM-based Software Engineering Systems}


\author{Francisco Ortin}
\orcid{0000-0003-1199-8649}
\affiliation{%
	\institution{University of Oviedo}
	\city{Oviedo}
	\country{Spain}
}
\email{ortin@uniovi.es}
\affiliation{%
	\institution{Munster Technological University}
	\city{Cork}
	\country{Ireland}
}
\email{francisco.ortin@mtu.ie}



\begin{abstract}

Large language models (LLMs) are increasingly used for software engineering tasks, yet their stochastic behavior challenges the validity, reproducibility, and comparability of their evaluations. Conventional practices such as reporting a single output, an average score, best-of-$N$, or pass@$k$ performance can obscure variability and estimation uncertainty, potentially leading to misleading conclusions about system reliability. This article presents SafeLLM4SE, a practical methodology and reporting standard for statistically principled evaluation of LLM-based software engineering systems. Rather than treating generated outputs as deterministic artifacts, SafeLLM4SE treats them as realizations of a stochastic process and distinguishes quality, stability, and estimation uncertainty. It combines adaptive sampling with confidence intervals, distribution-aware statistical comparisons, effect sizes, and a minimum reporting standard covering model configuration, reproducibility, evaluation procedures, and resource usage. SafeLLM4SE is also provided as an \href{https://github.com/francisco-ortin/safellm4se}{open-source software package} available on \href{https://pypi.org/project/SafeLLM4SE/}{PyPI}, enabling researchers and practitioners to reproduce and extend the methodology. We illustrate its application by comparing two LLMs on HumanEval, a benchmark of programming problems assessed through functional tests.

\end{abstract}




\maketitle

\section{Introduction}
\label{section:introduction}

Large language models (LLMs) are rapidly becoming part of the software engineering (SE) toolchain. They are now used across a broad range of SE activities, including code generation, test generation, program repair, refactoring, documentation, code review, and software maintenance~\cite{Hou2024}. LLMs increasingly support developers in automating a diverse set of SE tasks and workflows, including tasks of growing complexity~\cite{Chen2021}.

However, LLM-based systems differ from many traditional SE tools in an important respect: their outputs are stochastic. Given the same prompt and nominal model configuration, repeated invocations may produce different outputs and, consequently, different SE outcomes. For example, an LLM may generate a program that passes all tests
in one invocation and an implementation that fails the tests
in another invocation for exactly the same task. Thus, the outcome of an evaluation is not necessarily a fixed property of the system, but rather an observation of an underlying stochastic process.

Ouyang et al.~\cite{Ouyang2025} investigated non-determinism in ChatGPT code generation across 829 problems. They found substantial variation across repeated generations and, importantly, showed that setting the temperature to zero does not guarantee deterministic code generation. Therefore, LLM non-determinism can affect both software correctness and scientific conclusions, making it a methodological concern rather than merely an implementation detail~\cite{Ouyang2025}. This problem is especially acute in agentic SE systems, whose multi-step trajectories compound sources of randomness, yet are commonly evaluated using a single run per task, thereby obscuring execution-to-execution variability~\cite{Mustahsan2025}.

The stochastic nature of LLMs also creates challenges for reproducibility. The same prompts may not produce the same outputs even under the same nominal LLM configuration~\cite{Ouyang2025}, which complicates reproducibility in contexts such as scientific research and safety-critical applications. Moreover, LLM services may change over time because of model updates or modifications to their serving infrastructure, even when the same model identifier and nominal configuration are used~\cite{Chauvin2026,Kang2026}. Reproducible evaluation therefore requires reporting not only the experimental setup, but also how stochastic outputs were sampled and evaluated~\cite{Pineau2021}.

Despite this non-deterministic behavior, many researchers and practitioners continue to characterize LLM performance using a single output, the average performance across multiple outputs, best-of-$N$ (i.e., the best output among $N$ executions), or pass@$k$ (i.e., an estimate of the probability that at least one of $k$ generated outputs is correct). These metrics capture different aspects of model performance, but they may provide an incomplete picture of the underlying stochastic behavior.

To illustrate the practical consequences, Table~\ref{table:example-data} presents hypothetical HumanEval results in which different models are evaluated 100 times on the same set of code-generation problems, with functional correctness assessed through testing. For each evaluation, the success rate is computed as the proportion
of problems for which the generated solution passes all functional tests.

A single-output evaluation identifies \emph{A} as the best-performing model, whereas its average performance is lower than that of the other two models. Table~\ref{table:example-data} also illustrates how average performance estimates may depend on the number of samples. For example, \emph{C} achieves an average success rate of 0.36 at $N=10$ and 0.54 at $N=100$. Likewise, \emph{B} and \emph{C} show comparable average performance for $N=100$, but the variability of \emph{C} (SD = 0.344) is substantially greater than that of \emph{B} (SD = 0.175). Consequently, average performance alone does not fully characterize the behavior of either model. In addition, although \emph{C}'s average success rate is higher than \emph{B}'s, its 95\% confidence interval (CI) is substantially wider. This indicates that the estimated performance of \emph{C} is less precise, making comparisons and decisions based on the point estimate alone potentially less reliable.

Best-of-$N$ reports the best observed result within a sampling budget, whereas pass@$k$ estimates the probability of obtaining at least one correct result among $k$ samples. Neither alone characterizes the distribution of performance across complete executions. In the example, model \emph{C} achieves noticeably higher best-of-10 and pass@5 results than the other models, even though its average behavior does not show a comparable advantage. Therefore, these metrics should not be used as standalone measures when the goal is to characterize the overall stochastic behavior of LLM-based SE systems.

\begin{table*}
\caption{Hypothetical HumanEval results illustrating how sample size, variability, and uncertainty affect assessments of stochastic LLM-based systems.}
\label{table:example-data}
	\begin{tabular*}{\linewidth}{@{\extracolsep{\fill}} 
			>{\centering\arraybackslash}p{0.9cm} 
			>{\centering\arraybackslash}p{1.25cm} 
			>{\centering\arraybackslash}p{0.9cm} 
			>{\centering\arraybackslash}p{0.9cm} 
			>{\centering\arraybackslash}p{0.9cm} 
			>{\centering\arraybackslash}p{0.9cm} 
			>{\centering\arraybackslash}p{2.05cm} 
			>{\centering\arraybackslash}p{1.0cm} 
			>{\centering\arraybackslash}p{1.2cm} 
		}		\toprule
		\textbf{Model} & \textbf{Single Output} & \textbf{Average (N=10)} & \textbf{Average (N=30)} & \textbf{Average (N=100)} & \textbf{SD (N=100)}   & \textbf{95\% CI (N=100)} & \textbf{Pass@5}  &  \textbf{Best-of-10}  \\
		\midrule
		\emph{A}     &   0.658     &    0.277	& 0.268	 & 0.328 & 0.152 & [0.298,0.358] & 0.864 & 0.931	\\
		\emph{B}     &    0.424      &  0.509	& 0.542	& 0.529	& 0.175 & [0.495, 0.563] & 0.903 & 0.962 \\
		\emph{C}     &    0.425     &   0.360	&0.438 &	0.540	& 0.344 & [0.473,0.607] & 0.948 & 0.981 \\
		\bottomrule
	\end{tabular*}
\end{table*}

These limitations motivate a shift from point-based to distribution-based evaluation of LLMs for SE. The output of an LLM should be regarded as a sample from a stochastic process rather than as a deterministic artifact. Accordingly, the object of evaluation is not a single output, but the underlying distribution of possible outputs and their associated performance. This perspective makes it possible to distinguish three complementary properties of an LLM-based SE system: quality (expected performance of the system), stability (how much the system's performance varies across repeated executions), and estimation uncertainty (how precisely those quantities have been estimated from a finite number of observations).

To address these issues, we propose SafeLLM4SE, a practical methodology for the statistically principled evaluation and reporting of stochastic LLM-based SE systems. SafeLLM4SE is not intended to introduce a new statistical test. Rather, it defines a minimum standard for evaluation and reporting by integrating established statistical techniques into a workflow designed for SE researchers and practitioners. The framework provides default recommendations supported by an \href{https://github.com/francisco-ortin/safellm4se}{open-source software application}, while allowing justified alternatives when the characteristics of a particular experiment warrant them.

SafeLLM4SE is organized around five principles:

\begin{enumerate}
	
	\item Outputs are samples, not artifacts. For a fixed prompt, model, and configuration, each invocation is treated as an observation from a stochastic process. Evaluation therefore begins by identifying the performance metric induced by the generated artifact (e.g., pass/fail, BLEU score, or code quality).
	
	\item Distributions matter more than points. Repeated executions are used to characterize the distribution of performance rather than merely to obtain a point estimate such as an average. The framework distinguishes expected quality from variability in system performance.
	
	\item Inference should replace informal aggregation. SafeLLM4SE recommends quantifying estimation uncertainty using CIs, applying statistical tests appropriate to the experimental design, and reporting effect sizes rather than relying solely on differences in means or $p$-values.
	
	\item Quality, stability, and uncertainty are complementary. A reliable evaluation should report not only how well a system performs on average (quality), but also how consistently it performs across executions (stability) and how precisely these quantities have been estimated (uncertainty).
	
	\item Statistical rigor should be balanced against evaluation cost. Repeated LLM invocations incur API costs, latency, and computational resources. SafeLLM4SE therefore recommends adaptive sampling: an experiment begins with an initial number of executions and continues sampling until the estimate reaches a predefined precision target or the available budget is exhausted. This replaces arbitrary rules such as ``always run the benchmark 100 times'' with an explicit relationship between statistical precision and evaluation cost.
\end{enumerate}

Finally, we demonstrate the application of SafeLLM4SE by using its
Python implementation to evaluate and compare two LLMs on the
HumanEval benchmark.

The remainder of this article is organized as follows. Section~\ref{section:related-work} reviews prior work on stochastic evaluation, statistical analysis, and reproducible reporting, and situates SafeLLM4SE within these research directions. Section~\ref{section:SafeLLM4SE} presents the methodology, from probabilistic modeling and adaptive sampling to statistical characterization, reporting, comparison, and evidence-based decision making. Section~\ref{section:case-study} illustrates the methodology through a HumanEval case study, describing the experimental setup and showing how quality, stability, and estimation uncertainty inform the comparison of two LLMs. Finally, Section~\ref{section:conclusion} summarizes the main lessons and their implications for the evaluation of LLM-based SE systems.

\section{Related Work}
\label{section:related-work}

A growing body of empirical research documents non-determinism
in code generation by LLM-based systems. Beyond the study by Ouyang et al.~\cite{Ouyang2025} discussed in Section~\ref{section:introduction}, HumanEval itself was introduced together with the pass@$k$ estimator to summarize functional correctness over repeated samples~\cite{Chen2021}; however, pass@$k$ summarizes the probability of obtaining at least
one correct solution among $k$ samples per problem. Although pass@1
estimates the mean success probability for a single generation,
these point estimates alone do not characterize variability across
complete benchmark executions or quantify estimation uncertainty,
both of which are central to SafeLLM4SE. More broadly, Liang et al.~\cite{Liang2023} proposed HELM, a large-scale, multi-metric evaluation framework for language models that emphasizes standardization and transparency across scenarios. However, HELM does not prescribe how repeated stochastic generations should be sampled, aggregated, or statistically compared.

Concerns about treating machine learning benchmark results as fixed numbers rather than as outcomes of a stochastic process predate LLMs. Bouthillier et al.~\cite{Bouthillier2021} showed that multiple sources of randomness, beyond the commonly reported seed, can substantially affect benchmark outcomes and proposed variance-aware evaluation procedures. In deep reinforcement learning, Colas et al.~\cite{Colas2018} applied power analysis to determine adequate numbers of runs, and Agarwal et al.~\cite{Agarwal2021} advocated reporting interval estimates and robust aggregate statistics instead of isolated point estimates across runs. SafeLLM4SE builds on these ideas but adapts them to the specific setting of LLM-based SE systems, where evaluation cost is dominated by API usage and token consumption rather than training time, motivating its adaptive, budget-aware sampling procedure.

Within empirical SE, Arcuri and Briand~\cite{Arcuri2014} established influential guidelines for statistically analyzing randomized algorithms, recommending non-parametric tests, effect sizes, and a minimum number of repeated runs; SafeLLM4SE follows a similar spirit but targets LLM-based systems specifically, adding confidence-interval-driven adaptive sampling and a dedicated reporting standard. In the natural language processing community, Dror et al.~\cite{Dror2018} and Dodge et al.~\cite{Dodge2019} similarly criticized the reliance on single-run comparisons and proposed protocols for significance testing and computation-aware reporting, respectively, though neither targets the SE-specific artifacts and cost trade-offs (e.g., functional correctness, token budgets) that SafeLLM4SE addresses.

Finally, reporting checklists for reproducible machine learning research, such as the recommendations of Pineau et al.~\cite{Pineau2021}, share SafeLLM4SE's goal of making empirical results more transparent and comparable. SafeLLM4SE complements these general checklists with an SE-specific methodology that integrates adaptive sampling, distribution-aware statistical comparison, and a minimum reporting standard tailored to stochastic LLM-based SE evaluation. This methodology is made available through an \href{https://github.com/francisco-ortin/safellm4se}{open-source software tool} to lower the barrier to adoption~\cite{Ortin2026SafeLLM4SE}.

\section{SafeLLM4SE}
\label{section:SafeLLM4SE}

SafeLLM4SE defines a minimum evaluation protocol and reporting standard for experimental evaluations of LLM-based SE systems under stochastic generation. Rather than treating an LLM output as a deterministic artifact, SafeLLM4SE models it as a realization of a stochastic process and specifies the statistical evidence that should be reported to support reliable, reproducible, and comparable experimental conclusions.

\subsection{Probabilistic Modeling}
\label{subsection:SafeLLM4SE:probabilistic-modelling}

Given a prompt $p$, model $m$, and LLM configuration $c$, each execution of an LLM produces one realization of a random variable $Y_{p,m,c}$. Repeated executions under the same experimental conditions are treated as realizations from the underlying stochastic process governing the system's outputs.

Depending on the task and the chosen unit of analysis, each output or execution is evaluated using a metric $M(Y)$. Examples include functional correctness of generated code (pass or fail), the number of tests passed (integer-valued), execution time, or code quality (real-valued). Let $X_{p,m,c}=M(Y_{p,m,c})$ denote the resulting random variable. The evaluation is then based on samples of $X_{p,m,c}$ and aims to characterize its underlying performance distribution $F_{p,m,c}$.

SafeLLM4SE considers a target property $\theta$ of $F_{p,m,c}$, such as its mean or a success probability, and estimates it from repeated observations. The choice of $\theta$ depends on the metric and unit of analysis used in the evaluation. This distribution captures the expected performance of the system and its variability across repeated executions, whereas uncertainty arises from estimating properties of this distribution from a finite sample. Consequently, a single observation or point estimate such as the sample mean provides only a partial characterization of system behavior. SafeLLM4SE therefore treats repeated executions and the statistical characterization of the resulting distribution as fundamental components of LLM-based SE evaluation.

\subsection{Adaptive Sampling}
\label{subsection:SafeLLM4SE:adaptive-sampling}

Instead of fixing an arbitrary number of executions, SafeLLM4SE uses adaptive sampling to determine when a sufficient number of executions has been collected to estimate $\theta$, a target property of the performance distribution $F_{p,m,c}$, with the desired precision. This avoids both undersampling, which may lead to unreliable estimates, and unnecessary executions once additional samples provide little improvement in precision. Consequently, adaptive sampling can reduce API and computational costs, as well as the associated energy consumption and carbon footprint, while maintaining a predefined level of statistical precision.

For a given prompt $p$, model $m$, and configuration $c$, the procedure takes three parameters: a minimum sample size $N_{\min}$, a target confidence-interval precision $w_{\max}$, and a token-budget threshold $B$ expressed as a number of tokens. The observations collected during sampling are, as mentioned, realizations of $X_{p,m,c}=M(Y_{p,m,c})$. Algorithm~\ref{alg:adaptive-sampling} summarizes the procedure.

\begin{algorithm}
	\caption{Adaptive sampling}
	\label{alg:adaptive-sampling}
	\begin{algorithmic}[1]
		
		\REQUIRE Prompt $p$, model $m$, configuration $c$, minimum sample size $N_{\min}$, target precision $w_{\max}$, execution budget $B$ (in tokens) 
		\ENSURE Sample $S$, estimate $\hat{\theta}$ and tokens $T$
		
		\STATE Execute $(p,m,c)$ $N_{\min}$ times, collecting the corresponding observations of $X_{p,m,c}$ in $S$
		\STATE  $T \gets$total number of tokens consumed so far
		\STATE $n \gets |S|$
		\STATE Estimate $\theta$ from $S$ and obtain $\hat{\theta}$
		\STATE Compute a confidence interval $CI(S)$ for $\theta$
		
		\WHILE{$T < B$ \AND $\mathrm{width}(CI(S)) > w_{\max}$}
		\STATE Execute $(p,m,c)$ once and add the resulting observation of $X_{p,m,c}$ to $S$
		\STATE $T \gets T +$ tokens consumed by this call
		\STATE $n \gets n + 1$
		\STATE Estimate $\theta$ from $S$ and obtain $\hat{\theta}$
		\STATE Compute a confidence interval $CI(S)$ for $\theta$
		\ENDWHILE
		
		\RETURN $S$, $\hat{\theta}$, $T$
	\end{algorithmic}
\end{algorithm}

The confidence interval $CI$ is computed according to the type of metric being estimated. For continuous metrics, SafeLLM4SE uses a standard $t$-based CI when the sampled values can reasonably be assumed to follow an approximately normal distribution~\cite{Hollander2013}. When this assumption is not appropriate, SafeLLM4SE uses a bootstrap CI instead~\cite{Efron1993}. This provides a consistent procedure for metrics whose distributions may be skewed or otherwise depart substantially from normality. For binary metrics, such as whether generated code passes the
functional tests, SafeLLM4SE uses the Wilson score
interval~\cite{Wilson1927}. This is preferred to the standard
Wald interval, whose coverage can be poor when the estimated
proportion is close to~0 or~1.

The target precision $w_{\max}$ is defined as the maximum total width of the CI for the target property $\theta$. For example, a target width of 0.10 corresponds to a margin of error of $\pm 0.05$. Thus, the required number of executions is determined by the observed variability of $X_{p,m,c}$ rather than by an arbitrary fixed value: stable evaluation conditions may reach the target precision with relatively few executions, whereas highly variable conditions require additional samples. This allows SafeLLM4SE to adapt the number of executions to the observed variability of each evaluation instance while targeting a predefined CI width.

\subsection{Statistical Characterization}
\label{subsection:SafeLLM4SE:characterization}

SafeLLM4SE requires each evaluation to characterize the three dimensions of quality, stability, and estimation uncertainty, while allowing the specific indicators used for quality and stability to be selected according to the evaluation context. Together, these dimensions provide a more complete description of the performance distribution $F_{p,m,c}$ than any single point estimate.

\textbf{Quality} describes the central performance of the technique under the evaluation conditions. For a target property $\theta$ of $F_{p,m,c}$, SafeLLM4SE recommends reporting the arithmetic mean as the default measure of central tendency. When the distribution is clearly skewed or contains strong outliers, the median may be reported as a robust measure of typical performance. For binary outcomes, such as whether generated code compiles or whether all functional tests pass, quality is characterized by the probability of success, estimated as the proportion of successful executions among all executions. For proportion-valued outcomes, such as the benchmark success rate obtained in a complete benchmark execution, quality is characterized by the mean of that performance measure across repeated executions. The direction of the metric should also be made explicit (i.e., whether higher or lower values indicate better performance).

\textbf{Stability} describes the variability of $X_{p,m,c}$ across repeated executions under the same evaluation conditions. SafeLLM4SE recommends reporting the standard deviation as the default measure of dispersion. The coefficient of variation may additionally be reported when comparing techniques with different mean performance, provided that the metric has a meaningful ratio scale and a non-zero mean. For highly skewed or heavy-tailed distributions, the interquartile range provides a more robust alternative. Additional distributional descriptors, such as percentiles, may be reported when they provide useful information about the behavior of the system.

\textbf{Estimation uncertainty} arises because $F_{p,m,c}$ is inferred from a finite sample of executions. Consequently, the estimated property $\hat{\theta}$ is subject to sampling uncertainty. SafeLLM4SE therefore recommends reporting a CI for all primary performance indicators, using the procedure described in Section~\ref{subsection:SafeLLM4SE:adaptive-sampling}, with a 95\% confidence level by default. This uncertainty should be clearly distinguished from the variability of the LLM: stability describes how much $X_{p,m,c}$ varies across executions, whereas the CI describes the precision with which the target property $\theta$ of $F_{p,m,c}$ has been estimated from the available sample.

\subsection{Reporting And Visualization}
\label{subsection:SafeLLM4SE:reporting}

SafeLLM4SE defines the minimum information that an evaluation should report to allow readers to assess the statistical validity, reproducibility, and practical reliability of its results. The reporting requirements cover the three dimensions mentioned—quality, stability, and estimation uncertainty—as well as the experimental protocol used to generate and evaluate the observations.

Table~\ref{table:report} summarizes the SafeLLM4SE reporting standard. For each evaluated technique, researchers should report the selected quality and stability indicators and their corresponding estimates. CIs and the confidence level used to construct them are required for all primary performance indicators.

\begin{table*}
\caption{SafeLLM4SE minimum reporting standard for evaluations of LLM-based SE techniques.}
\label{table:report}
	\begin{tabular*}{\linewidth}{@{\extracolsep\fill} L{1.8cm} L{7.6cm} L{4.8cm}}
		\toprule
		\textbf{Category}                & \textbf{Information to report}                                                                                                            & \textbf{Purpose}                                                        \\
		\midrule
		Quality                 & Selected quality indicator(s) and estimated value $\hat{\theta}$.                                 & Characterize the performance of the technique.                  \\
		Stability               & Selected stability indicator(s), such as standard deviation,   coefficient of variation, or interquartile range.                  & Characterize variability across repeated executions.            \\
		Estimation \newline uncertainty  & CI for each primary performance indicator and confidence level.                                                & Quantify the precision of the reported estimates.               \\
		Sampling                & Number of executions, sampling procedure,   minimum sample size, stopping criterion, and execution budget, if applicable.         & Assess the adequacy and efficiency of the sampling procedure.   \\
		Model                   & Model name, model version or identifier, and provider.                                                                            & Identify the system being evaluated and facilitate replication. \\
		Inference \newline configuration & Temperature and other inference parameters that may affect generation.                                                            & Specify the conditions under which outputs were generated.      \\
		Reproducibility         & Random seed, when supported; exact prompt(s); execution date and time;   and, when applicable, API or model snapshot information. & Account for sources of variation and support reproducibility.   \\
		Evaluation              & Benchmark or task version, evaluation procedure, test suite or other   evaluation artifacts, and software/tool versions.          & Make the evaluation procedure reproducible.                     \\
		Resources               & Number of generated tokens or equivalent usage information and, when   available, execution cost.                                 & Quantify the computational and economic resources required.    \\
		\bottomrule
	\end{tabular*} 
\end{table*}

For each estimated quality indicator $\hat{\theta}$, the reported estimate should be accompanied by its CI. Reporting only a point estimate, such as the mean success rate, is therefore insufficient because it does not indicate the precision of the estimate. Likewise, reporting a CI without describing the number of executions and the sampling procedure makes it difficult to assess how the estimate was obtained.

The reporting of model and execution information is particularly important for evaluations using commercial LLM APIs. Model behavior may change over time because of model updates or changes to the serving infrastructure, even when the same model identifier is used. Researchers should therefore report the exact model identifier or version, the execution date and time, and any available snapshot or deployment information. Random seeds should also be reported when the provider exposes them, although a seed should not be assumed to guarantee reproducibility unless the underlying service provides such a guarantee.

Whenever possible, numerical summaries should be complemented with visualizations of the observed values of $X_{p,m,c}$. Boxplots, violin plots, raincloud plots, or empirical cumulative distribution functions (ECDFs) can reveal differences in variability, skewness, heavy tails, and other distributional characteristics that may be obscured by summary statistics alone. The HumanEval case study illustrates this approach (Figure~\ref{figure:case-study-distribution}).

Collectively, these requirements constitute the SafeLLM4SE reporting standard. They provide reviewers and practitioners with the information needed to distinguish the observed performance of an LLM-based SE system from its variability and estimation uncertainty, while providing sufficient experimental detail to support meaningful replication and comparison.

\subsection{Statistical Comparison}
\label{subsection:SafeLLM4SE:comparison}

Comparisons between LLM-based SE techniques should be performed over the observed values of $X_{p,m,c}$ and the corresponding performance distributions $F_{p,m,c}$, rather than over isolated point estimates. An essential distinction is whether the observations are independent or paired. Samples are \emph{independent} when there is no one-to-one correspondence between observations from the two techniques, such as when two techniques are evaluated using different sets of users or independently selected software projects. Samples are \emph{paired} when both techniques are evaluated under the same experimental conditions, creating a natural correspondence between observations. For example, when two models are evaluated on the same HumanEval problems, the results for each problem form a pair. Pairing applies to problem-level outcomes; complete benchmark executions are paired only if the experimental design links individual runs across techniques.

SafeLLM4SE defines a default statistical procedure for each experimental design, summarized in Table~\ref{table:comparison}. These procedures constitute the recommended minimum protocol rather than a collection of interchangeable alternatives.

\begin{table}
\caption{SafeLLM4SE statistical comparison protocol.}
\label{table:comparison}
	\begin{tabular*}{\linewidth}{@{\extracolsep\fill} L{1.8cm} L{3.4cm} L{4.2cm} L{4.8cm}}
		\toprule
		\textbf{Design} & \textbf{Significance test} & \textbf{Confidence interval (CI)} & \textbf{Effect size} \\
		\midrule
		Independent &
		Mann--Whitney U &
		Bootstrap CI for the \newline difference &
		Cliff's $\delta$ \\
		Paired &
		Wilcoxon signed-rank &
		Paired bootstrap CI &
		Matched-pairs rank-biserial correlation \\
		\bottomrule
	\end{tabular*} 
\end{table}

For independent samples, SafeLLM4SE specifies the Mann--Whitney U test because it provides a non-parametric comparison of two distributions without requiring normally distributed observations~\cite{Hollander2013}. The test is complemented by Cliff's $\delta$, which describes how often values from one technique exceed those from the other~\cite{Cliff1993}. The difference in quality, $\theta_A-\theta_B$, is estimated by
$\hat{\theta}_A-\hat{\theta}_B$. A bootstrap CI for $\theta_A-\theta_B$
quantifies uncertainty in this estimate without requiring a
parametric distributional assumption.

For paired samples with suitably symmetric within-pair differences, SafeLLM4SE recommends the Wilcoxon signed-rank test on the per-instance differences $\hat{\theta}_{p,A,c}-\hat{\theta}_{p,B,c}$~\cite{Hollander2013}. For markedly asymmetric differences or binary paired outcomes, the test should be chosen according to the data. The matched-pairs rank-biserial correlation is reported as the effect size because it quantifies the balance between favorable and unfavorable signed ranks and provides a directional measure of the magnitude of the paired effect~\cite{Kerby2014}. A paired bootstrap CI for $\theta_A-\theta_B$ is computed by
resampling complete pairs, thereby preserving the dependence
structure between the two techniques.

The three reported quantities provide complementary information. The statistical test indicates whether the observed data provide evidence of a difference according to the selected test statistic, the CI quantifies the precision of the estimated difference in the target property, and the effect size describes the magnitude of the distributional or paired effect. Statistical
significance should therefore not be interpreted as evidence of practical importance on its own: with sufficiently large samples, even very small differences may become statistically significant.

SafeLLM4SE also recommends reporting the magnitude of non-parametric effect sizes using established interpretation guidelines. For independent samples, Cliff's $\delta$ ranges from $-1$ to $1$
and measures the difference between the probability that an
observation from one technique exceeds an observation from the
other and the reverse probability~\cite{Cliff1993}. Values close
to zero indicate that these probabilities are approximately
balanced, but do not necessarily imply similar distributions.
Larger absolute values indicate a stronger tendency for one
technique to yield higher values than the other. The conventional thresholds are approximately $|\delta|=0.147$, $0.33$, and $0.474$ for small, medium, and large effects, respectively. For paired comparisons, the matched-pairs rank-biserial correlation has the same $[-1,1]$ range and provides an analogous interpretation of effect direction and magnitude~\cite{Kerby2014}. These thresholds should be regarded as general guidelines rather than universal definitions of practical importance; the substantive relevance of an effect ultimately depends on the task, metric, and application context.

Alternative statistical procedures may be used when the experimental design or structure of the data makes the default protocol inappropriate or violates its underlying assumptions. Such deviations should be explicitly justified and reported. Regardless of the procedure used, a statistical comparison should
report the estimated difference $\hat{\theta}_A-\hat{\theta}_B$,
its CI, the statistical test used, the resulting $p$-value,
the chosen significance level, and the effect size. Together, these quantities allow readers to assess whether a difference is statistically detectable, how precisely it has been estimated, and whether its magnitude is practically meaningful.

\subsection{Decision Making}
\label{subsection:SafeLLM4SE:decision-making}

SafeLLM4SE has two complementary objectives: (1) to provide a statistically principled methodology for analyzing stochastic LLM-based SE systems, and (2) to establish a minimum reporting standard that facilitates reproducibility, comparability, and evidence-based decision making across LLM4SE studies.

The evidence collected through SafeLLM4SE should support, rather than replace, experimental judgment. Let $\theta_A$ and $\theta_B$ denote the target properties of the performance distributions $F_A$ and $F_B$ for two competing techniques under the same evaluation conditions, and let $\hat{\theta}_A$ and $\hat{\theta}_B$ be their estimates. A higher observed value $\hat{\theta}_A$ alone is therefore not sufficient to conclude that technique $A$ is superior to technique $B$. Such a conclusion should consider the estimated difference, its uncertainty, the magnitude of the effect, and the stability of the two techniques.

A claim that one technique is superior to another should be supported by evidence that considers:

\begin{enumerate}
	
	\item Quality. The technique achieves a better value of the target
property $\theta$, taking into account the direction of the metric.
For example, a higher success probability indicates better quality
for binary correctness outcomes, whereas a lower mean execution time
indicates better performance for latency measurements.
	
	\item Estimation uncertainty. The observed difference between $\hat{\theta}_A$ and $\hat{\theta}_B$ should be interpreted together with its CI. A difference should not be considered conclusive when the available data provide insufficient precision to distinguish the two techniques reliably (i.e., when the CI for the estimated difference is compatible with both a meaningful advantage and no meaningful difference).	
	
	\item Statistical evidence. The observed difference should be supported by the statistical analysis described in Section~\ref{subsection:SafeLLM4SE:comparison}, rather than being plausibly attributable to sampling variation alone.
	
	\item Practical relevance. The estimated effect should be sufficiently large to be meaningful for the task and application, rather than being statistically significant but practically negligible.
	
	\item Stability. The variability of $X_{p,m,c}$ should be considered alongside quality and estimation uncertainty. Lower variability generally indicates more stable performance, but greater variability may be an acceptable trade-off when it is accompanied by a sufficiently meaningful improvement in quality.
	
\end{enumerate}

These criteria should be considered jointly. A technique may, for example, achieve a higher $\hat{\theta}$ while also exhibiting greater variability or substantial estimation uncertainty. In such cases, declaring a clear overall winner may be inappropriate, and the corresponding trade-offs should be reported explicitly. Similarly, when the estimated difference is small,
imprecisely estimated, or associated with a negligible effect size, researchers should avoid definitive superiority claims.

SafeLLM4SE therefore does not prescribe a universal ranking rule or a single decision criterion. Instead, it provides the statistical evidence needed to distinguish clear improvements from trade-offs and from differences that cannot be reliably established. This supports transparent and evidence-based decisions about which LLM-based SE technique is preferable for a given task and application context.

\section{Case Study: Applying SafeLLM4SE To HumanEval}
\label{section:case-study}

This section illustrates the application of SafeLLM4SE to the evaluation and comparison of two LLMs. We consider \texttt{gemini-3.1-flash-lite}, accessed through the Gemini API, and \texttt{qwen2.5-coder:7b}, accessed through Ollama. Both models are evaluated on the 164 problems of HumanEval using a temperature of 2.0. The experiment deliberately uses a relatively high temperature to amplify stochastic variation and make its effects more apparent in the evaluation results.

\subsection{Experimental Setup}
\label{subsection:case-study:setup}

We treat HumanEval as a single benchmark evaluation comprising its 164 programming problems. Each execution of a model consists of generating one solution for each of the 164 problems and evaluating the generated solutions using the HumanEval functional tests. Thus, each complete execution produces one observation of the model's stochastic performance on the benchmark.

For each model, we apply the adaptive sampling proposed by SafeLLM4SE. We use $N_{\min}=30$, an unlimited execution budget, and require the 95\% bootstrap CI for the mean benchmark success rate to achieve a maximum total width of $w_{\max} = 0.01$ (i.e., a maximum margin of error of $\pm 0.005$). Since no budget limit is imposed in this case study, sampling stops
once at least $N_{\min}$ complete benchmark executions have been
collected and the observed CI width is at or below $w_{\max}$.

The outcome for each generated solution is binary: $X_{p,m,c,i}=1$ when the solution generated for problem $p$ in execution $i$ passes all HumanEval tests, and $X_{p,m,c,i}=0$ otherwise. Thus, the performance of execution $i$ is summarized as the proportion of correctly solved HumanEval problems,

\begin{equation}
	X_{m,c,i}
	=
	\frac{1}{164}
	\sum_{p=1}^{164} X_{p,m,c,i}.
\end{equation}

Thus, $X_{m,c,i}$ is a proportion-valued performance measure in the interval $[0,1]$, with higher values indicating better quality. The target property for quality, $\theta_{m,c}$, is the expected
benchmark success rate under the evaluated conditions. We estimate
this quantity using the sample mean of the success rates observed
across $N$ complete benchmark executions:

\begin{equation}
	\hat{\theta}_{m,c}
	=
	\frac{1}{N}
	\sum_{i=1}^{N} X_{m,c,i}.
\end{equation}

Quality is therefore estimated from the distribution of complete benchmark executions rather than from separate problem-level success probabilities.

For each model, stability is assessed from the variability of $X_{m,c,i}$ across repeated executions. We report the standard deviation of the benchmark success rate as the primary stability measure. 

The CI for the mean benchmark success rate is computed from the $N$ execution-level observations using bootstrap resampling. Each bootstrap sample consists of $N$ complete benchmark executions sampled with replacement, and the mean success rate is computed for each sample. The 95\% percentile bootstrap CI is defined by the 2.5th and
97.5th percentiles of the bootstrap distribution of the mean
benchmark success rate. This interval quantifies uncertainty in the estimated mean performance
across repeated executions of the fixed set of 164 HumanEval problems.

\subsection{SafeLLM4SE Reporting}
\label{subsection:case-study:reporting}

Table~\ref{table:case-study-reporting} reports the experimental results according to the SafeLLM4SE reporting standard defined in Table~\ref{table:report}. The results show that Gemini achieves a higher mean benchmark success rate, $\hat{\theta}_{m,c}$, of 0.9335 compared with 0.8238 for Qwen-Coder. Gemini also exhibits a lower observed standard deviation across complete HumanEval executions (0.0110 vs.\ 0.0204), indicating lower observed variability under the evaluated conditions.

\begin{table*}
	\caption{HumanEval results reported according to the SafeLLM4SE reporting standard.}
	\label{table:case-study-reporting}
	\small
	\begin{tabular*}{\linewidth}{@{\extracolsep\fill} L{2.1cm} L{4.3cm} L{4.3cm} L{4.3cm}}
		\toprule
		\textbf{Category} & \textbf{Information} &
		\textbf{Gemini} & \textbf{Qwen-Coder} \\
		\midrule
		Quality &
		Mean benchmark success rate ($\hat{\theta}_{m,c}$)  &
		0.9335 & 0.8238 \\
		
		Stability &
		Standard deviation of benchmark success rate &
		0.0110 & 0.0204 \\
		
		Estimation \newline uncertainty  &
		95\% CI for mean benchmark success rate &
		[0.9294, 0.9376] & [0.8188, 0.8287] \\
		
		Sampling &
		$N_{\min}$; target CI; budget; $N$; total generated solutions &
		30; width of 95\% CI $\leq 0.01$; unlimited; 30; 4,920 &
		30; width of 95\% CI $\leq 0.01$; unlimited; 67; 10,988  \\
		
		Model &
		Model name, identifier, and provider &
		Gemini, \texttt{gemini-\discretionary{}{}{}3.1-\discretionary{}{}{}flash-\discretionary{}{}{}lite}; Gemini API v1beta&
		Qwen-Coder, \texttt{qwen2.5-\discretionary{}{}{}co\-der\-:\-7b}; Ollama v0.34.4 \\
		
		Inference \newline configuration &
		Temperature &
		2.0 & 2.0 \\
		
		Reproducibility &
		Execution date; Prompts; API/model information; seed &
		Sep. 24, 2026, 18:46; HumanEval v0.1.10 prompts as available on that date; online Gemini API as available on that date; seed unavailable &
		Sep. 24, 2026, 18:46; HumanEval v0.1.10 prompts as available on that date; Ollama v0.34.4; seed unavailable \\
		
		Evaluation &
		Benchmark and evaluation procedure &
		HumanEval (164 problems); functional tests &
		HumanEval (164 problems); functional tests \\
		
		Resources &
		Number of generated tokens and benchmark executions &
		2,542,131; 30 &
		2,635,240; 67 \\
		\bottomrule
	\end{tabular*}
\end{table*}

\begin{figure*}
	\centering
	\includegraphics[width=0.9\linewidth]{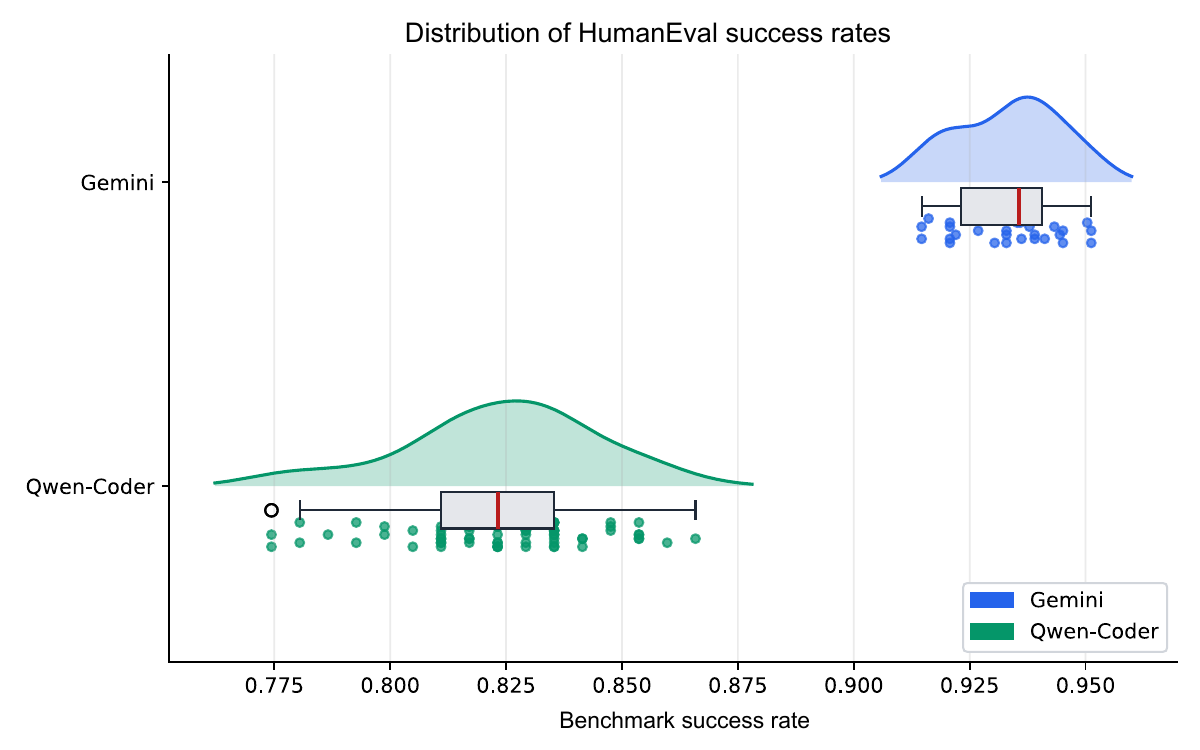}
	\caption{Distribution of benchmark success rates of Gemini and Qwen-Coder across repeated HumanEval executions obtained with SafeLLM4SE adaptive sampling. Each observation represents one complete execution of the 164 HumanEval problems. Figure generated with the SafeLLM4SE Python tool.}
	\label{figure:case-study-distribution}
\end{figure*}

Figure~\ref{figure:case-study-distribution} shows the distribution of benchmark success rates across complete HumanEval executions for both models. Each observation corresponds to one complete execution of the 164 problems, providing a direct view of stochastic variation in benchmark-level performance. Gemini's success rates are concentrated at higher values than those of Qwen-Coder and exhibit lower dispersion across repeated executions, indicating greater observed stability.

\subsection{Statistical Comparison}
\label{subsection:case-study:comparison}

The two models are evaluated through repeated complete executions of the same HumanEval benchmark, with each execution comprising the same 164 benchmark problems. The execution-level benchmark success rate $X_{m,c,i}$ is used to characterize the quality and stability of each model across repeated executions. For the statistical comparison between models, the two samples of execution-level success rates are treated as independent: each execution is an unlinked realization of the model's own stochastic process, with no run-level correspondence between models.

Following the SafeLLM4SE comparison protocol for independent observations, we compare the two samples of execution-level benchmark success rates using the Mann--Whitney U test. Cliff's $\delta$ is reported as the effect size. In addition, we obtain a bootstrap CI for the difference in mean benchmark success rate by independently resampling the execution-level observations of each model with replacement, without assuming any correspondence between individual executions of the two models. The results are shown in Table~\ref{table:case-study-comparison}.

\begin{table}
\caption{SafeLLM4SE comparison of the two models.}
\label{table:case-study-comparison}
	\begin{tabular*}{.9\linewidth}{@{\extracolsep\fill} L{10cm} L{4cm}}
		\toprule
		\textbf{Comparison} & \textbf{Result} \\
		\midrule
		Mean difference &
		$+0.1097$ \\
		95\% bootstrap CI for the difference in mean success rate &
		$[0.1033, 0.1160]$ \\
		Mann--Whitney U test &
		$U=2010$,  $\ p<0.001$ \\
		Cliff's $\delta$ &
		$1.0$ \\
		\bottomrule
	\end{tabular*}
\end{table}

\subsection{Decision Making}
\label{subsection:case-study:decision-making}

The results provide consistent evidence in favor of Gemini's \texttt{gemini-3.1-flash-lite} under the evaluated conditions. Its mean benchmark success rate is 10.97 percentage points higher than that of Qwen-Coder, with a 95\% CI for the difference entirely above zero. Gemini also exhibits lower observed variability across repeated HumanEval executions. The Mann--Whitney U test indicates a difference between the observed distributions, and Cliff's $\delta=1$ indicates complete stochastic separation in the observed samples: every observed Gemini benchmark score exceeds every observed Qwen-Coder score. Moreover, Gemini used fewer total tokens to reach the target CI width in this experiment. Thus, Gemini satisfies the SafeLLM4SE criteria for concluding that it provides superior performance under the evaluated conditions and exhibits lower observed variability across repeated benchmark executions. 

This case study was conducted entirely using the SafeLLM4SE application, developed as part of this work and used to generate all the data and Figure~\ref{figure:case-study-distribution}. The application is freely available for download~\cite{Ortin2026SafeLLM4SE} and is published on PyPI under the name \href{https://pypi.org/project/SafeLLM4SE/}{SafeLLM4SE}.

\section{Conclusions}
\label{section:conclusion}

The main lesson from this work is that trustworthy evaluation of LLM-based SE systems requires treating their outputs as observations of a stochastic process rather than as deterministic artifacts. A single generation or point estimate can conceal substantial variability and may therefore provide an insufficient basis for engineering decisions. Trustworthy evaluation should instead characterize three complementary dimensions: quality, stability, and estimation uncertainty. This perspective shifts the focus from asking which system produced the highest observed score in a particular evaluation to asking what performance can be expected across repeated executions and how precisely that performance has been estimated.

A second lesson is that statistical rigor and practical evaluation constraints should be considered together. Repeated executions are necessary to characterize stochastic behavior, but fixed sample sizes can either provide insufficient evidence or incur unnecessary cost. Adaptive sampling relates the number of executions to a predefined CI-width target, concentrating evaluation effort where additional observations are needed. Its stopping rule and interval coverage should be reported and assessed together. Similarly, comparisons should consider uncertainty and effect magnitude, rather than relying on statistical significance or differences in point estimates alone. These principles make the resulting evidence more useful for informed engineering decisions and reduce the risk of drawing strong conclusions from inherently variable systems.

Ultimately, trustworthy AI-enabled SE depends not only on the capabilities of the underlying models, but also on the quality and transparency of the evidence used to evaluate them. SafeLLM4SE provides a practical basis for producing such evidence and for making evaluations more reproducible, comparable, and accountable. 

The software developed in this work, together with its source code and the data used in the case study, is freely available~\cite{Ortin2026SafeLLM4SE}.

\begin{acks}
	
This work was supported by the Ministry of Science, Innovation and Universities and co-funded by the European Regional Development Fund (ERDF) under project PID2024-155586OB-I00 (MCIU/AEI/10.13039 /501100011033). Additional support was provided by the Government of the Principality of Asturias through grant GRU-GIC-24-070.

\end{acks} 

\bibliographystyle{ACM-Reference-Format}
\bibliography{bibliography}

@article{Hou2024,
	author  = {Xinyi Hou and Yanjie Zhao and Yue Liu and Zhou Yang and
	Kailong Wang and Li Li and Xiapu Luo and David Lo and
	John Grundy and Haoyu Wang},
	title   = {Large Language Models for Software Engineering:
	A Systematic Literature Review},
	journal = {ACM Transactions on Software Engineering and Methodology},
	volume  = {33},
	number  = {8},
	pages   = {1--79},
	year    = {2024},
	doi     = {10.1145/3695988}
}

@article{Ouyang2025,
	author  = {Shuyin Ouyang and Jie M. Zhang and Mark Harman and Meng Wang},
	title   = {An Empirical Study of the Non-Determinism of {ChatGPT}
	in Code Generation},
	journal = {ACM Transactions on Software Engineering and Methodology},
	volume  = {34},
	number  = {2},
	pages   = {1--28},
	year    = {2025},
	doi     = {10.1145/3697010}
}

@article{Chen2021,
	author  = {Mark Chen and Jerry Tworek and Heewoo Jun and Qiming Yuan
	and Henrique Ponde de Oliveira Pinto and Jared Kaplan and
	Harri Edwards and Yuri Burda and Nicholas Joseph and
	Greg Brockman and Alex Ray and Raul Puri and Gretchen Krueger
	and Michael Petrov and Heidy Khlaaf and Girish Sastry and
	Pamela Mishkin and Brooke Chan and Scott Gray and Nick Ryder
	and Mikhail Pavlov and Alethea Power and Lukasz Kaiser and
	Mohammad Bavarian and Clemens Winter and Philippe Tillet and
	Felipe Petroski Such and Dave Cummings and Matthias Plappert and
	Fotios Chantzis and Elizabeth Barnes and Ariel Herbert-Voss and
	William Hebgen Guss and Alex Nichol and Alex Paino and Nikolas Tezak
	and Jie Tang and Igor Babuschkin and Suchir Balaji and Shantanu Jain
	and William Saunders and Christopher Hesse and Andrew N. Carr and
	Jan Leike and Josh Achiam and Vedant Misra and Evan Morikawa and
	Alec Radford and Matthew Knight and Mira Murati and Katie Mayer and
	Peter Welinder and Bob McGrew and Dario Amodei and
	Sam McCandlish and Ilya Sutskever and Wojciech Zaremba},
	title   = {Evaluating Large Language Models Trained on Code},
	journal = {arXiv preprint arXiv:2107.03374},
	year    = {2021}
}

@article{Pineau2021,
	author  = {Joelle Pineau and Philippe Vincent-Lamarre and Koustuv Sinha and
	Vincent Larivi{\`e}re and Alina Beygelzimer and Florence
	d'Alch{\'e}-Buc and Emily Fox and Hugo Larochelle},
	title   = {Improving Reproducibility in Machine Learning Research
	(A Report from the {NeurIPS} 2019 Reproducibility Program)},
	journal = {Journal of Machine Learning Research},
	volume  = {22},
	number  = {164},
	pages   = {1--20},
	year    = {2021}
}

@inproceedings{Chauvin2026,
	author    = {Timoth{\'e}e Chauvin and Erwan Le Merrer and
	Fran{\c{c}}ois Ta{\"i}ani and Gilles Tredan},
	title     = {Log Probability Tracking of {LLM APIs}},
	booktitle = {Proceedings of the International Conference on Learning
	Representations},
	year      = {2026}
}

@inproceedings{Kang2026,
	author    = {Daniel Kang and others},
	title     = {Behavioral Fingerprints for {LLM} Endpoint Stability and Identity},
	booktitle = {Proceedings of the ACM Conference on AI and Agentic Systems},
	pages     = {1327--1331},
	year      = {2026},
	doi       = {10.1145/3786335.3813194}
}

@article{Cliff1993,
	author  = {Cliff, Norman},
	title   = {Dominance Statistics: Ordinal Analyses to Answer Ordinal Questions},
	journal = {Psychological Bulletin},
	volume  = {114},
	number  = {3},
	pages   = {494--509},
	year    = {1993},
	doi     = {10.1037/0033-2909.114.3.494}
}

@article{Kerby2014,
	author  = {Kerby, Dave S.},
	title   = {The Simple Difference Formula: An Approach to Teaching Nonparametric Correlation},
	journal = {Comprehensive Psychology},
	volume  = {3},
	pages   = {1},
	year    = {2014},
	doi     = {10.2466/11.IT.3.1}
}

@misc{Ortin2026SafeLLM4SE,
	author       = {Ortin, Francisco},
	title        = {{SafeLLM4SE}},
	year         = {2026},
	url          = {https://github.com/francisco-ortin/safellm4se},
	note         = {{GitHub} repository},
	accessed     = {2026-10-07}
}

@book{Efron1993,
	author    = {Efron, Bradley and Tibshirani, Robert J.},
	title     = {An Introduction to the Bootstrap},
	publisher = {Chapman \& Hall/CRC},
	year      = {1993},
	series    = {Monographs on Statistics and Applied Probability},
	volume    = {57},
	address   = {Boca Raton, FL, USA}
}

@article{Wilson1927,
	author  = {Wilson, Edwin B.},
	title   = {Probable Inference, the Law of Randomness, and Single Samples},
	journal = {Journal of the American Statistical Association},
	volume  = {22},
	number  = {158},
	pages   = {209--212},
	year    = {1927},
	doi     = {10.1080/01621459.1927.10502953}
}

@article{Liang2023,
	author  = {Liang, Percy and Bommasani, Rishi and Lee, Tony and Tsipras, Dimitris and Soylu, Dilara and Yasunaga, Michihiro and Zhang, Yian and Narayanan, Deepak and Wu, Yuhuai and Kumar, Ananya and Newman, Benjamin and Yuan, Binhang and Yan, Bobby and Zhang, Ce and Cosgrove, Christian and Manning, Christopher D. and R{\'e}, Christopher and Acosta-Navas, Diana and Hudson, Drew A. and Zelikman, Eric and Durmus, Esin and Ladhak, Faisal and Rong, Frieda and Ren, Hongyu and Yao, Huaxiu and Wang, Jue and Santhanam, Keshav and Orr, Laurel and Zheng, Lucia and Yuksekgonul, Mert and Suzgun, Mirac and Kim, Nathan and Guha, Neel and Chatterji, Niladri and Khattab, Omar and Henderson, Peter and Huang, Qian and Chi, Ryan and Xie, Sang Michael and Santurkar, Shibani and Ganguli, Surya and Hashimoto, Tatsunori and Icard, Thomas and Zhang, Tianyi and Chaudhary, Vishrav and Wang, William and Li, Xuechen and Mai, Yifan and Zhang, Yuhui and Koreeda, Yuta},
	title   = {Holistic Evaluation of Language Models},
	journal = {Transactions on Machine Learning Research},
	year    = {2023}
}

@inproceedings{Bouthillier2021,
	author    = {Bouthillier, Xavier and Delaunay, Pierre and Bronzi, Mirko and Trofimov, Assya and Nichyporuk, Brennan and Szeto, Justin and Sepahvand, Nazanin Mohammadi and Raff, Edward and Madan, Kanika and Voleti, Vikram and others},
	title     = {Accounting for Variance in Machine Learning Benchmarks},
	booktitle = {Proceedings of Machine Learning and Systems},
	volume    = {3},
	pages     = {747--769},
	year      = {2021}
}

@article{Colas2018,
	title     = {How Many Random Seeds? {S}tatistical Power Analysis in Deep Reinforcement Learning Experiments},
	author    = {Colas, C{\'e}dric and Sigaud, Olivier and Oudeyer, Pierre-Yves},
	journal   = {arXiv preprint arXiv:1806.08295},
	year      = {2018},
	eprint    = {1806.08295},
	archivePrefix = {arXiv},
	primaryClass  = {cs.LG},
	url       = {https://arxiv.org/abs/1806.08295}
}

@article{Agarwal2021,
	author  = {Agarwal, Rishabh and Schwarzer, Max and Castro, Pablo Samuel and Courville, Aaron and Bellemare, Marc G.},
	title   = {Deep Reinforcement Learning at the Edge of the Statistical Precipice},
	journal = {arXiv preprint arXiv:2108.13264},
	year    = {2021}
}

@article{Arcuri2014,
	author  = {Arcuri, Andrea and Briand, Lionel},
	title   = {A Hitchhiker's Guide to Statistical Tests for Assessing Randomized Algorithms in Software Engineering},
	journal = {Software Testing, Verification and Reliability},
	volume  = {24},
	number  = {3},
	pages   = {219--250},
	year    = {2014}
}

@inproceedings{Dror2018,
	author    = {Dror, Rotem and Baumer, Gili and Shlomov, Segev and Reichart, Roi},
	title     = {The Hitchhiker's Guide to Testing Statistical Significance in Natural Language Processing},
	booktitle = {Proceedings of the 56th Annual Meeting of the Association for Computational Linguistics (Volume 1: Long Papers)},
	pages     = {1383--1392},
	year      = {2018}
}

@inproceedings{Dodge2019,
	author    = {Dodge, Jesse and Gururangan, Suchin and Card, Dallas and Schwartz, Roy and Smith, Noah A.},
	title     = {Show Your Work: Improved Reporting of Experimental Results},
	booktitle = {Proceedings of the 2019 Conference on Empirical Methods in Natural Language Processing and the 9th International Joint Conference on Natural Language Processing (EMNLP-IJCNLP)},
	pages     = {2185--2194},
	year      = {2019}
}

@misc{Mustahsan2025,
	title={Stochasticity in Agentic Evaluations: Quantifying Inconsistency with Intraclass Correlation}, 
	author={Zairah Mustahsan and Abel Lim and Megna Anand and Saahil Jain and Bryan McCann},
	year={2025},
	eprint={2512.06710},
	archivePrefix={arXiv},
	primaryClass={cs.AI},
	url={https://arxiv.org/abs/2512.06710}, 
}

@book{Hollander2013,
	author    = {Hollander, Myles and Wolfe, Douglas A. and Chicken, Eric},
	title     = {Nonparametric Statistical Methods},
	edition   = {3rd},
	publisher = {John Wiley \& Sons},
	address   = {Hoboken, NJ},
	year      = {2013},
	isbn      = {978-0-470-38737-5}
}

\end{document}